\documentclass[conference,a4paper]{IEEEtran}
\IEEEoverridecommandlockouts
\usepackage{cite}
\usepackage{amsmath,amssymb,amsfonts}
\usepackage{algorithmic}
\usepackage{graphicx}
\usepackage{textcomp}
\usepackage{xcolor}
\def\BibTeX{{\rm B\kern-.05em{\sc i\kern-.025em b}\kern-.08em
    T\kern-.1667em\lower.7ex\hbox{E}\kern-.125emX}}

\newcommand*{\cbra}[1]{\left\{ #1 \right\}}
\newcommand*{\rbra}[1]{\left( #1 \right)}

\newcommand*{\set}[2]{\left\{\, #1 \;\middle|\; #2 \,\right\}}
\newcommand*{\abs}[1]{\left\lvert #1 \right\rvert}

\DeclareMathOperator*{\argmin}{arg\,min}

\renewcommand{\epsilon}{\varepsilon}
\renewcommand{\theta}{\vartheta}
\renewcommand{\kappa}{\varkappa}
\renewcommand{\rho}{\varrho}
\renewcommand{\phi}{\varphi}

\usepackage{eso-pic} 
\usepackage{multirow} 
\usepackage{url} 
\usepackage{cleveref}

\begin{document}

\title{Scalable quantile predictions of peak loads for non-residential customer segments
\thanks{This publication is part of the project ROBUST: Trustworthy AI-based Systems for Sustainable Growth with project number KICH3.LTP.20.006, which is partly financed by the Dutch Research Council (NWO).}
}

\author{\IEEEauthorblockN{Shaohong Shi}
\IEEEauthorblockA{\textit{Dept.~of Electrical Sustainable Energy} \\
\textit{Delft University of Technology}\\
Delft, the Netherlands \\
s.shi-1@tudelft.nl
}
\and
\IEEEauthorblockN{Jacco Heres}
\IEEEauthorblockA{\textit{Research Centre for Digital Technologies} \\
\textit{Alliander N.V.}\\
Arnhem, the Netherlands \\
jacco.heres@alliander.com
}
\and
\IEEEauthorblockN{Simon H.~Tindemans}
\IEEEauthorblockA{\textit{Dept.~of Electrical Sustainable Energy} \\
\textit{Delft University of Technology}\\
Delft, the Netherlands \\
s.h.tindemans@tudelft.nl
}
}

\maketitle


\begin{abstract}
Electrical grid congestion has emerged as an immense challenge in Europe, making the forecasting of load and its associated metrics increasingly crucial. Among these metrics, peak load is fundamental. Non-time-resolved models of peak load have their advantages of being simple and compact, and among them Velander's formula (VF) is widely used in distribution network planning. However, several aspects of VF remain inadequately addressed, including year-ahead prediction, scaling of customers, aggregation, and, most importantly, the lack of probabilistic elements. The present paper proposes a quantile interpretation of VF that enables VF to learn truncated cumulative distribution functions of peak loads with multiple quantile regression under non-crossing constraints. The evaluations on non-residential customer data confirmed its ability to predict peak load year ahead, to fit customers with a wide range of electricity consumptions, and to model aggregations of customers. A noteworthy finding is that for a given electricity consumption, aggregations of customers have statistically larger peak loads than a single customer.
\end{abstract}


\section{Introduction}

Electrical grid congestion has become a significant issue in Europe. The cost incurred by EU transmission system operators for remedial actions to alleviate physical grid congestion is projected to rise from EUR 4.26 billion in 2023 \cite{EUACER2024} to EUR 11\textendash  26 billion in 2030, and EUR 34\textendash  103 billion in 2040 \cite{Thomassen2024}. In order to manage congestion and reduce costs, it is essential to predict loads and their relevant metrics. 

Load forecasting models predict load profiles \cite{Khuntia2016}, from which metrics such as peak load, lost load, and the value of lost load can be derived. 
However, for the peak load, whose accuracy improvement brings the greatest economic benefit \cite{Ranaweera1997}, non-time-resolved methods that model peak loads directly are usually simpler and more compact. Examples include Velander's formula (VF) \cite{Velander1935}, Rusck's diversity factor \cite{Rusck1956}, and the simple form customer class load model \cite{Seppaelae1996}. Their simplicity renders them more practical and easier for DSOs to deploy. Furthermore, they are indispensable in modeling peak loads for customers without smart meter data and for new customers.

Among non-time-resolved methods, VF is well established. It was first used by Scandinavian distribution system operators (DSOs), and was adopted by DSOs in the Netherlands and other European countries thanks to \cite{Axelsson1975}. Case studies have shown its efficacy in providing reliable estimates of peak loads for individual customers \cite{Persson2018} and aggregations \cite{Velander1935,Velander1952}. This is partly due to its relationship with Rusck's coincidence factor \cite{Oirsouw2011}, which is beneficial in scenarios such as connecting new customers and dispatching substations. 

Nevertheless, four issues about VF remain to be addressed. First, although parameters of VF fitted on historical data are often used for prediction by DSOs, there is little literature quantifying its practicability. On the contrary, a case study of a residential district shows that new parameters need to be derived for a winter with abnormal weather conditions \cite{Braennlund2011}. This casts doubt on the aforementioned applicability.

The second issue pertains to the scaling of peak load with electricity consumption (EC). In contrast to households, ECs and peak loads of large customers usually span a broad range, even in one segment (see Appendix 7 of \cite{Persson2018}). However, we have not found any literature investigating whether VF represents the relationship between peak loads and ECs of both smaller individuals and larger individuals in one segment at the same time.

The third issue relates to the validity for aggregation of customers. The authors of \cite{Fuerst2020} claimed an improvement in the accuracy of modeling aggregated peak loads by implementing an additional parameter in VF to capture correlation between customers, which suggests possible flaws regarding aggregated peak loads in VF.

Last but not least, VF is a deterministic model, whilst it is increasingly important for DSOs to model the uncertainty in power loads \cite{Heres2017}. Much research has been dedicated to probabilistic models of load profiles \cite{Khajeh2022}. Additionally, there is also some research on probabilistic non-time-resolved models. Examples include investigating probability distribution of Rusck's diversity factor \cite{Nazarko1998,Chatlani2007}, peak load using fuzzy regression \cite{Nazarko1999} and quantile regression (QR) \cite{Gibbons2014}, and probability distribution of peak load with extreme value theory \cite{Lee2022}. Incorporating a ``probabilistic VF'' is therefore a compelling prospect, especially if the advantages of VF can be retained. 

\textbf{Contribution of this paper.} The paper proposes a quantile interpretation of VF that enables VF to learn truncated cumulative distribution functions (CDFs) of peak loads with multiple quantile regression (MQR) under non-crossing constraints. The aforementioned issues were evaluated, including its potential for year-ahead prediction, its ability to fit customers with a broad range of ECs, and its effectiveness in modeling aggregated peak loads. The experiments yielded positive results for the first two issues, while the third issue led to the discovery of a rightward shift in the CDFs of aggregated peak loads (corresponding to the same aggregated EC) as the aggregation level increases.

\section{Quantile Velander's Formula}

We introduce VF and our quantile interpretation of VF. Let \(T \in \mathbb{N}_{>0}\), and denote by \([T]\) the discrete time range \(\cbra{1, 2, \ldots, T}\), where each interval lasts \(\Delta\). Let \(C\) be an index set, and let \(\set{(P_i(t))_{t \in [T]}}{i \in C}\) be a set of load profiles. Define the peak load \(P_{i,\mathrm{max}}\) and the EC \(E_i\) of \(i \in C\) by
\[P_{i,\mathrm{max}} := \max_{t \in [T]} P_i(t), \quad E_i := \sum_{t \in [T]} P_i(t) \Delta.\]
Furthermore, let \(\mathcal{E}\) be the set \(\set{E_i}{i \in C}\), and let \(\mathcal{E}_a\) be the \(a\)th percentile of \(\mathcal{E}\) for \(a \in [0,100]\).

Velander stated in \cite{Velander1935} that, if the load profiles are of the same type, there exist parameters \(\alpha\) and \(\beta\) corresponding to that type, such that the predicted peak load \(\hat{P}_{i,\mathrm{max}}\) is given by \begin{equation} \label{eq-Velander} \hat{P}_{i,\mathrm{max}} := \alpha E_i + \beta \sqrt{E_i}, \quad i \in C, \end{equation}
In \cite{Velander1952} Velander fitted \eqref{eq-Velander} for loads of power plants and cities. In practice, \(\alpha\) and \(\beta\) can be fitted with peak loads and ECs of individual customers \cite{Braennlund2011}, of aggregations of randomly chosen customers \cite{Persson2018}, or even of mixture of individuals and aggregations \cite{Oirsouw2011}.
Then \eqref{eq-Velander} is used to estimate aggregated peak loads by regarding aggregations as a virtual customer of the same type. 

However, it is not obvious how \eqref{eq-Velander} should be understood or applied to different situations. The authors of \cite{Fuerst2020} assumed \(\set{P_i(t)}{t \in T}\) to be a set of independent and identically distributed (i.i.d.) Gaussian samples, and regarded \(\hat{P}_{i, \mathrm{max}}\) as the \(\tau\)-quantile of the Gaussian distribution for a fixed \(\tau \in (0,1)\). That assumption seldom holds in practice, and the \(\tau\)-quantile has no direct relation to the peak load.

Instead, we regard \(\hat{P}_{i,\mathrm{max}}\) in \eqref{eq-Velander} as the \(\tau\)-quantile \(P^{(\tau)}_{i,\mathrm{max}}\) of \(P_{i,\mathrm{max}}\), a direct probabilistic metric on peak load:
\begin{equation} \label{eq-Velander-quantile} P^{(\tau)}_{i,\mathrm{max}} = \alpha_\tau E_i + \beta_\tau \sqrt{E_i}, \quad \tau \in (0,1), i \in C, \end{equation}
where \(\alpha_\tau\) and \(\beta_\tau\) are functions of \(\tau\). The \(C\) in \eqref{eq-Velander-quantile} refers only to a set of individual customers, while aggregations will be discussed later in \Cref{subsection-a13.50}.

\section{Multiple Quantile Regression} \label{section-mqr}

MQR is an extension of QR, whose loss function is the summation of pinball losses at different quantile levels plus an optional regularizer \cite{Takeuchi2006}. It is extended from the pinball loss, and was proved to be a proper scoring rule in \cite{Cervera1996}.

Let \(A\) be a finite set of distinct quantile levels, and denote by \(\theta\) the parameter set \(\set{(\alpha_\tau,\beta_\tau)}{\tau \in A}\). Let \(\Theta\) be a parameter space of \(\theta\). The MQR for \eqref{eq-Velander-quantile} is given by 
\begin{equation} \label{eq-MQR} \theta := \argmin_{\theta \in \Theta} \mathrm{APL}(C, [T], \theta),\end{equation}
where the APL, the \emph{average pinball loss}, is defined by
\begin{equation}\label{eq-MQR-APL} \mathrm{APL}(C, [T], \theta) := \frac{1}{|C|} \frac{1}{|A|} \sum_{i \in C} \sum_{\tau \in A} \mathrm{PL}(i,\tau),\end{equation}
\begin{equation*} \mathrm{PL}(i,\tau) := \begin{cases} (\tau-1)   \delta(i,\tau)  & \text{if } \delta(i,\tau) < 0 \\
\tau \delta(i,\tau) & \text{if } \delta(i,\tau) \ge 0, \end{cases}\end{equation*}
\[\delta(i,\tau) := P_{i,\mathrm{max}} - \alpha_\tau E_i - \beta_\tau \sqrt{E}_i.\]

When no constraint is imposed in \eqref{eq-MQR} (\textbf{C1}), the MQR degenerates into separate QRs for \(\tau \in A\). However, estimated quantile functions for different quantile levels may cross or overlap \cite{He1997}. MQR is capable of avoiding quantile crossings with non-crossing constraints. We define three constraints for \eqref{eq-MQR} in ascending order of strictness as follows.

\par{\textbf{C2}. The avoidance of crossings means that, for \(x \in \mathcal{E}\),
\begin{equation*} \alpha_{\tau_1} x + \beta_{\tau_1} \sqrt{x} \le \alpha_{\tau_2} x + \beta_{\tau_2} \sqrt{x}, \quad \tau_1 \le \tau_2, \;\; \tau_1, \tau_2 \in A,\end{equation*} 
or, equivalently, for \(x \in \mathcal{E}\),
\begin{equation} \label{eq-C2} (\alpha_{\tau_1} - \alpha_{\tau_2}) \sqrt{x} + \beta_{\tau_1}-\beta_{\tau_2} \le 0, \quad \tau_1 \le \tau_2, \;\; \tau_1, \tau_2 \in A.\end{equation}
}

\par{\textbf{C3}. A natural tighter version of \eqref{eq-C2} is to extend \(x \in \mathcal{E}\) to \(x > 0\), which is then equivalent to 
\begin{equation} \label{eq-C3} \alpha_{\tau_1} \le \alpha_{\tau_2}, \quad \beta_{\tau_1} \le \beta_{\tau_2}, \quad \tau_1 \le \tau_2, \;\; \tau_1, \tau_2 \in A.\end{equation}
}

\par{\textbf{C4}. When performing the MQR with real data, \(\alpha_\tau\) was found to be always almost constant across \(\tau \in A\), which inspired us to adapt \eqref{eq-C3} to 
\begin{equation} \label{eq-C4} \alpha_{\tau_1} = \alpha_{\tau_2}\equiv \alpha, \quad \beta_{\tau_1} \le \beta_{\tau_2}, \quad \tau_1 \le \tau_2, \;\; \tau_1, \tau_2 \in A.\end{equation}
}

\section{Analyses}

We introduce how we evaluate \eqref{eq-Velander-quantile} in terms of non-crossing constraints, year-ahead prediction, scaling and aggregation.

\subsection{Non-crossing constraints} \label{subsection-non-crossing-constraints}

Under each constraint, a 5-fold cross-validation is performed to obtain the average training and testing APLs of the MQR, which is used to compare the constraints.

To provide an intuition of the quantity of the APLs, for \(i \in C\), a synthetic load profile is generated by sampling \(T\) i.i.d.\ loads from the Gaussian distribution \(N(\mu_i, \sigma_i^2)\), where \[\mu_i := T^{-1} E_i, \quad \sigma_i^2 := T^{-1} \sum_{t \in [T]} (P_i(t) - \mu_i)^2.\]
Similarly, a 5-fold cross-validation is performed under constraint C1 on the synthetic load profiles. Although real load profiles do not consist of i.i.d.\ Gaussian samples as the synthetic ones, they have similar distribution of ECs of customers, and their corresponding APLs should be comparable.

\subsection{Year-ahead prediction} \label{subsection-year-ahead-prediction}

Denote by \(\mathcal{T}_1\) and \(\mathcal{T}_2\) two successive discrete time ranges with length \(T\), and denote by \(C_1\) and \(C_2\) the corresponding set of customers in the same segment in \(\mathcal{T}_1\) and \(\mathcal{T}_2\), respectively. Define the parameter sets \(\theta(\mathcal{T}_1)\) and \(\theta(\mathcal{T}_2)\) by 
\[\theta(\mathcal{T}_i) := \argmin_{\theta \in \Theta} \mathrm{APL}(C_i, \mathcal{T}_i, \theta), \quad i \in {1,2}.\]
To evaluate the time invariance of the parameters in \eqref{eq-Velander-quantile}, we define the \emph{temporal loss difference} (TLD) by 
\begin{equation} \label{eq-TLD} \mathrm{TLD}(\mathcal{T}_2 | \mathcal{T}_1) := \frac{\mathrm{APL}(C_2, \mathcal{T}_2, \theta(\mathcal{T}_1))}{\mathrm{APL}(C_2, \mathcal{T}_2, \theta(\mathcal{T}_2))} -1.
\end{equation}
It shows how much the APL in \(\mathcal{T}_2\) with the parameter set trained on historical time range \(\mathcal{T}_1\) exceeds the minimal APL in \(\mathcal{T}_2\). The closer the TLD is to 0, the more consistent \(\theta\) is over time, and the better \eqref{eq-Velander-quantile} performs in year-ahead prediction.

\subsection{Scaling} \label{subsection-scaling}

We train the parameter set \(\theta\) on the smaller (larger) half of \(C\) and test it on the other half to evaluate whether \eqref{eq-Velander-quantile} is able to accommodate the scaling of customer sizes. For \(a, b \in [0,100]\) with \(a < b\), define the set \(C_{a,b}\) of customers by \[C_{a,b} := \set{i \in C}{\mathcal{E}_a \le E_i < \mathcal{E}_b}.\] 
If \(C_{a,b} \neq \emptyset\), we define the parameter set \(\theta(C_{a,b})\) by 
\[\theta(C_{a,b}) := \argmin_{\theta \in \Theta} \mathrm{APL}(C_{a,b}, [T], \theta).\]
If, in addition, \(c, d \in [0,100]\) and \(C_{c,d} \neq \emptyset\), we define the \emph{scaling loss difference} (SLD) analogous to the TLD by
\begin{equation} \label{eq-SLD} \mathrm{SLD} (C_{a,b} | C_{c,d}) := \frac{\mathrm{APL}(C_{a,b}, [T], \theta(C_{c,d}))}{\mathrm{APL}(C_{a,b}, [T], \theta(C_{a,b}))} - 1, 
\end{equation}
Here, we take \(((a, b), (c, d))\) to be \(((0, 50), (50, 100))\) and \(((50, 100),  (0, 50))\). Then the SLDs show the relative difference between the testing APL on the smaller (larger) customers with the parameter set fitted on the larger (smaller) customers and the training APL on the smaller (larger) customers. The closer the SLD is to zero, the better \eqref{eq-Velander-quantile} fits smaller and larger customers simultaneously.

\subsection{Aggregation} \label{subsection-a13.50}

For the customer subset \(C' \subseteq C\), let \(P_{C'}(t)\) be the aggregated load \(\sum_{i \in C'} P_i(t)\) at \(t \in [T]\). Define the aggregated peak load \(P_{C',\mathrm{max}}\) and the aggregated EC \(E_{C'}\) by
\[P_{C',\mathrm{max}} := \max_{t \in [T]} P_{C'}(t), \quad E_{C'} := \sum_{i \in C'} E_i,\] 
respectively. We call \(\abs{C'}\) the \emph{aggregation level} of the aggregation consisting of customers in \(C'\).

Let \(l \in \cbra{1, 2 ,\ldots, \abs{C}}\), and let \(\mathcal{S}_l\) be the set of subsets of \(l\) customers in \(C\). Analogous to \eqref{eq-Velander-quantile}, we propose
\begin{equation} \label{eq-Velander-quantile-aggr} P^{(\tau)}_{C_l,\mathrm{max}} = \alpha^{(l)}_\tau E_{C_l} + \beta^{(l)}_\tau \sqrt{E_{C_l}}, \quad \tau \in (0,1), C_l \in \mathcal{S}_l. \end{equation}
It would be appealing if an aggregation is equivalent to an individual customer with the same EC, namely, for \(\tau \in (0,1)\),
\begin{equation} \label{eq-Velander-quantile-aggr-assumption} \alpha^{(l)}_\tau = \alpha_\tau, \beta^{(l)}_\tau = \beta_\tau, \quad l \in \mathbb{N}_{>0}.\end{equation}

To evaluate \eqref{eq-Velander-quantile-aggr}, for \(l \in \{2, 5, 10, 25\}\), random groups of \(l\) customers are sampled and aggregated independently 1000 times. Then a five-fold cross-validation of the MQR is performed to obtain the average training and testing APL, which will be normalized, namely, divided by \(l\), for comparison among aggregations at different aggregation levels.

We assess \eqref{eq-Velander-quantile-aggr-assumption} by comparing QR curves and truncated CDFs of customers as follows. Let \(D_1\) be the set \[\set{(E_i, P_{i,\mathrm{max}})}{i \in C, \mathcal{E}_{40} \le E_i \le \mathcal{E}_{60}}.\]
Let \(\mathcal{S}'_2\) and \(\mathcal{S}'_3\) compromise \(4 |C|\) and \(16|C|\) random sampled elements from \(\mathcal{S}_2\) and \(\mathcal{S}_3\), respectively, and define \(D_l\) as \[\set{(E_{C_l}, P_{C_l, \mathrm{max}})}{C_l \in \mathcal{S}'_l, \mathcal{E}_{40} \le E_{C_l} \le \mathcal{E}_{60}}, \quad l \in \{2,3\}.\]
For \(l \in \{1,2,3\}\), the parameter set \(\set{(\alpha^{(l)}_\tau,\beta^{(l)}_\tau)}{\tau \in A}\) is obtained by minimizing the APL corresponding to \(D_l\). Here, we limit the ECs to \([\mathcal{E}_{40}, \mathcal{E}_{60}]\) to alleviate the effect of the wide range of ECs.

For \(l \in \{1,2,3\}\), we plot the QR curves with the function \[y = \alpha^{(l)}_\tau x + \beta^{(l)}_\tau \sqrt{x}\] for \(\tau \in \cbra{0.2,0.5,0.8}\), and the approximate truncated CDFs of peak loads of aggregations with EC \(E\) at aggregation level \(l\) with the points 
\[\rbra{\alpha^{(l)}_\tau E + \beta^{(l)}_\tau \sqrt{E}, \tau}, \quad \tau \in A,\] 
for \(E \in \{\mathcal{E}_{40}, \mathcal{E}_{50}, \mathcal{E}_{60}\}\). If \eqref{eq-Velander-quantile-aggr-assumption} holds, then the QR curves and CDFs for different aggregation levels should coincide.

\section{Data and Results}

Smart meter data of large customers (with a grid capacity between 60kW and 100MW) for the years 2022, 2023 and 2024 (abbreviated as 22, 23 and 24, respectively) were collected from Dutch DSO Liander with a 15-minute resolution. The units of loads and ECs are kW and kW\(\cdot\)15 minute, respectively. We performed analyses on three categories of large customers: SBI code 8411 (general government administration), SBI code 6420 (financial holding companies) and KvK code 004 (industry). \Cref{tabl-number-customers} shows the number of customers in the original data and the number of customers after removing customers whose load profiles are incomplete, have negative values, and are all zero at the first 672 time points (during the first week). We restrict \(\tau\) in \([0.1, 0.9]\) and fix the set \(A\) of quantile levels to be \(\cbra{0.10,0.11, \ldots, 0.90}\).

\begin{table}[htbp]
\caption{Number of customers (original \(\rightarrow\) processed)}
\begin{center}
\begin{tabular}{|c|c|c|c|}
\hline
\textbf{Year} & 2022 & 2023 & 2024 \\ \hline
SBI code 8411 & 1058 \(\rightarrow\) 778 & 1197\(\rightarrow\) 794 & 1202 \(\rightarrow\) 873 \\ \hline
SBI code 6420 & 1065 \(\rightarrow\) 709 & 1143 \(\rightarrow\) 727 & 1144 \(\rightarrow\) 765 \\ \hline
KvK code 004 & 1261 \(\rightarrow\) 952 & 1324 \(\rightarrow\) 934 & 1325 \(\rightarrow\) 961 \\ \hline
\end{tabular}
\label{tabl-number-customers}
\end{center}
\end{table}

\subsection{Non-crossing constraints} \label{subs-results-1-constraints}

\Cref{table-apls-constraints} shows the average training and testing APLs of the 5-fold cross-validation under each constraint, where ``C1*'' corresponds to MQRs performed under C1 on synthetic load profiles. Note that APLs on real data and on corresponding synthetic data are comparable, and the difference between average training APLs and their corresponding average test APLs is slight. Moreover, the difference among the test APLs under the four constraints is also very small, despite that there are 162 parameters in \eqref{eq-Velander-quantile} under C1, and only 82 parameters in \eqref{eq-Velander-quantile} under C4. Notably, among the nine average test APLs under C4, the six in bold font in \Cref{table-apls-constraints} are even smaller than their corresponding average test APLs under C1. This strongly indicates that the constraint C4, namely \eqref{eq-C4}, is the most suitable one for the large customers that we tested. Hence, \textbf{we applied C4 by default} in our following tests. 

\begin{table}[htbp]
\caption{Average training (Tr) and testing (Te) APLs (kW)}
\begin{center}
\begin{tabular}{|c|c|c||c|c||c|c|} \hline
Constraint & 22-Tr & 22-Te & 23-Tr & 23-Te & 24-Tr & 24-Te  \\ \hline
\multicolumn{7}{|c|}{SBI code 8411} \\ \hline
C1 & 30.75 & 31.28 & 28.43 & 30.91 & 27.87 & 28.15 \\ \hline
C2 & 30.74 & 31.38 & 28.45 & 30.59 & 27.83 & 28.62 \\ \hline
C3 & 30.71 & 31.84 & 28.46 & 30.48 & 27.91 & 28.01 \\ \hline
C4 & 30.83 & \bf{30.99} & 28.93 & \bf{29.30} & 27.92 & 28.27 \\ \hline
C1* & 19.89 & 20.66 & 23.06 & 24.13 & 17.18 & 17.93 \\ \hline
\multicolumn{7}{|c|}{SBI code 6420} \\ \hline
C1 & 23.56 & 26.48 & 24.11 & 25.06 & 28.65 & 29.53 \\ \hline
C2 & 23.33 & 28.78 & 23.96 & 26.48 & 28.69 & 29.17 \\ \hline
C3 & 23.44 & 28.02 & 23.95 & 26.63 & 28.41 & 32.48 \\ \hline
C4 & 24.79 & \bf{25.18} & 24.70 & 26.86 & 29.12 & 31.76 \\ \hline
C1* & 25.74 & 29.74 & 24.92 & 28.59 & 25.22 & 27.44 \\ \hline
\multicolumn{7}{|c|}{KvK code 004} \\ \hline
C1 & 42.15 & 44.01 & 42.26 & 44.68 & 40.84 & 42.29 \\ \hline
C2 & 42.24 & 43.18 & 42.34 & 43.91 & 40.80 & 42.70 \\ \hline
C3 & 42.12 & 44.50 & 42.33 & 44.24 & 40.82 & 42.43 \\ \hline
C4 & 42.97 & \bf{43.94} & 42.99 & \bf{43.62} & 41.27 & \bf{42.08} \\ \hline
C1* & 63.55 & 64.59 & 62.29 & 62.98 & 61.90 & 64.26 \\ \hline
\end{tabular}
\label{table-apls-constraints}
\end{center}
\end{table}

\subsection{Year-ahead prediction}

\Cref{table-TLD} shows the TLDs defined in \eqref{eq-TLD} for \((\mathcal{T}_2, \mathcal{T}_1)\) being the pair of years 2023 and 2022, and of years 2024 and 2023. The ECs in 2024 were multiplied by \(365/366\) in the calculation since 2024 is a leap year. The TLDs are very close to zero, which confirm the time invariance of parameters in \eqref{eq-Velander-quantile}, and provide a solid base for the utilization of \eqref{eq-Velander-quantile} in year-ahead prediction. 

\begin{table}[htbp]
\caption{Temporal loss differences}
\begin{center}
\begin{tabular}{|c|c|c|c|} \hline
Years & SBI 8411 & SBI 6420 & KvK 004 \\ \hline
(23,22) & 0.358\% & 0.619\% & 0.153\% \\ \hline
(24,23) & 0.523\% & 0.165\% & 0.258\% \\ \hline
\end{tabular}
\label{table-TLD}
\end{center}
\end{table}

\subsection{Scaling}

\begin{figure*}
\centering
\begin{minipage}[t]{0.32\linewidth}
    \includegraphics[width=\linewidth]{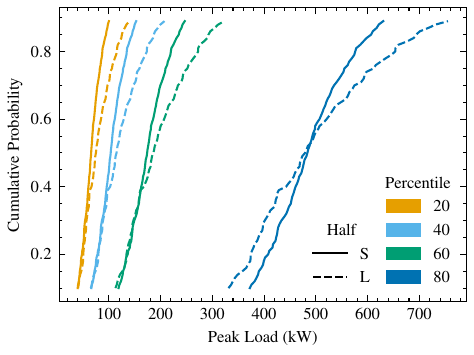}%
    \caption{KvK 004, 2024: CDFs of peak loads based on the smaller (S) and larger (L) halves.} \label{fig-VF-scaling-CDF-2024-KvK-004}
  \end{minipage}\hfill                 
\begin{minipage}[t]{0.32\linewidth}
    \includegraphics[width=\linewidth]{{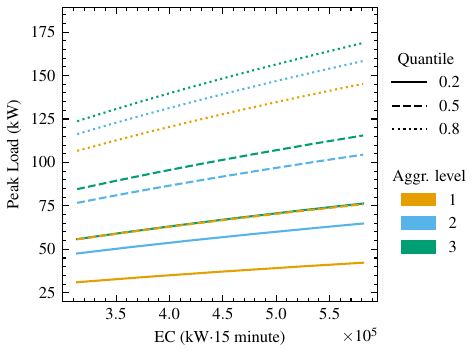}}
    \caption{SBI 8411, 2022: QR curves of peak loads of aggregations at aggregation levels \(1, 2\) and \(3\).}
    \label{fig-aggr-QR-2022-8411}
\end{minipage}\hfill
\begin{minipage}[t]{0.32\linewidth}
    \includegraphics[width=\linewidth]{{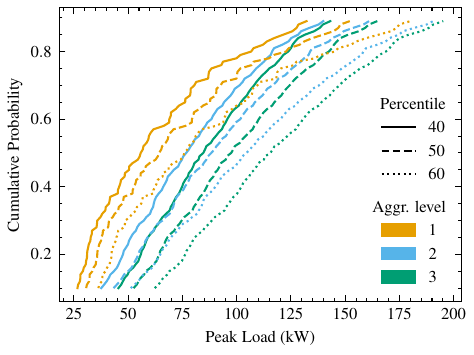}}
    \caption{SBI 8411, 2022: CDFs of peak loads of aggregations at aggregation levels \(1, 2\) and \(3\).} 
    \label{fig-aggr-CDF-2022-8411}
\end{minipage}
\end{figure*}

\Cref{table-tt-ratios} shows \(\mathrm{SLD}(C_{0,50} | C_{50,100})\) (labeled with ``S\textbar L'') and \(\mathrm{SLD}(C_{50, 100} | C_{0, 50})\) (labeled with ``L\textbar S''). To visualize the worst SLDs of KvK code 004 in year 2024, \Cref{fig-VF-scaling-CDF-2024-KvK-004} presents the truncated approximate CDFs of virtual customers with different percentiles of \(\mathcal{E}\) as their ECs. The results calculated with \(\theta(C_{0,50})\) and \(\theta(C_{50, 100})\) are plotted in solid and dotted lines, respectively. In general, the SLDs demonstrate the capability of \eqref{eq-Velander-quantile} to fit smaller and larger customers in the same segment with the same parameter values.

\begin{table}[htbp]
\caption{Scaling loss differences (in \%)}
\begin{center}
\begin{tabular}{|c||c|c||c|c||c|c|} \hline
\multirow{2}{*}{Category} & \multicolumn{2}{c||}{2022} & \multicolumn{2}{c||}{2023} & \multicolumn{2}{c|}{2024} \\ \cline{2-7}
 & S\textbar L & L\textbar S & S\textbar L & L\textbar S & S\textbar L & L\textbar S \\ \hline
SBI 8411 & 0.607 & 0.767 & 2.82 & \underline{17.8} & 2.63 & 1.28 \\ \hline
SBI 6420 & 3.53 & 3.65 & 2.75 & 4.20 & 2.74 & 1.24 \\ \hline
KvK 004 & 7.33 & 11.3 & 6.84 & 4.17 & 14.3 & \underline{22.6} \\ \hline
\end{tabular}
\label{table-tt-ratios}
\end{center}
\end{table}

We investigated the two anomalous values that are underlined in \Cref{table-tt-ratios}, which signify that the corresponding \(\theta(C_{0, 50})\) do not fit their respective \(C_{50, 100}\) well. The hypothesis that some very small customers were different from the rest was confirmed by the fact that the anomalous values returned to normal after their removal. For segment SBI code 8411 in 2023, \(\mathrm{SLD}(C_{3,50}, C_{50,100})\) and \(\mathrm{SLD}(C_{50, 100}, C_{3, 50})\) became 3.11\% and 7.55\%, respectively. For segment KvK code 004 in 2024, \(\mathrm{SLD}(C_{5,50}, C_{50,100})\) and \(\mathrm{SLD}(C_{50, 100}, C_{5, 50})\) became 10.7\% and 13.1\%, respectively.

Besides, the SLDs for the segment KvK code 004 are generally larger. This is possibly due to the more diverse customer behaviors within that segment, which comprises customers with multiple SBI codes.

\subsection{Aggregation}

\Cref{table-apls-aggregation} shows the average normalized training and testing APLs of the MQR at aggregation levels 2, 5, 10 and 25. Note that the difference between the training APLs and their corresponding testing APLs is slight. In addition, the average normalized APLs decrease as their aggregation levels increase, suggesting that \eqref{eq-Velander-quantile-aggr} fits better at larger aggregation levels.

\begin{table}[htbp]
\caption{Average normalized training (Tr) and testing (Te) APLs (kW) at different aggregation levels (A.L.)}
\begin{center}
\begin{tabular}{|c|c|c||c|c||c|c|} \hline
A.L. & 22-Tr & 22-Te & 23-Tr & 23-Te & 24-Tr & 24-Te  \\ \hline
\multicolumn{7}{|c|}{SBI code 8411} \\ \hline
2 & 23.55 & 23.67 & 26.64 & 27.24 & 21.15 & 21.19 \\ \hline
5 & 19.82 & 19.88 & 18.36 & 18.58 & 16.05 & 16.10 \\ \hline
10 & 15.58 & 15.61 & 13.95 & 14.10 & 13.22 & 13.24 \\ \hline
25 & 9.91 & 9.93 & 10.73 & 10.78 & 9.45 & 9.47 \\ \hline
\multicolumn{7}{|c|}{SBI code 6420} \\ \hline
2 & 20.61 & 21.08 & 20.20 & 21.24 & 26.40 & 26.77 \\ \hline
5 & 13.28 & 13.60 & 13.26 & 14.07 & 16.50 & 16.97 \\ \hline
10 & 9.52 & 9.80 & 10.46 & 10.61 & 13.87 & 14.46 \\ \hline
25 & 7.06 & 7.31 & 6.74 & 6.75 & 11.75 & 11.91 \\ \hline
\multicolumn{7}{|c|}{KvK code 004} \\ \hline
2 & 33.37 & 34.14 & 34.55 & 34.98 & 32.27 & 32.47 \\ \hline
5 & 25.84 & 25.95 & 25.29 & 25.49 & 23.17 & 23.33 \\ \hline
10 & 19.47 & 19.54 & 19.42 & 19.51 & 18.76 & 19.06 \\ \hline
25 & 12.43 & 12.56 & 13.33 & 13.39 & 12.66 & 12.69 \\ \hline
\end{tabular}
\label{table-apls-aggregation}
\end{center}
\end{table}

In addition, we found a systematic upward shift in the QR curves and rightward shift in the CDFs as the aggregation level \(l\) increases, as illustrated in \Cref{fig-aggr-QR-2022-8411,fig-aggr-CDF-2022-8411}, which suggests the failure of \eqref{eq-Velander-quantile-aggr-assumption} and that, for a given EC, aggregations of more customers have statistically larger peak loads.

\section{Conclusion}

The evaluations of the quantile interpretation of Velander's formula showed that \eqref{eq-Velander-quantile} performs extremely well in year-ahead prediction, and well for customers with a wide range of ECs on the three segments of large customers. In terms of aggregations, \eqref{eq-Velander-quantile-aggr} fits aggregated peak loads better as the aggregation level increases, while the assumption \eqref{eq-Velander-quantile-aggr-assumption} fails since the CDFs of aggregated peak loads corresponding to the same EC shifted rightward as the aggregation level increased.

Ongoing research will elucidate the mathematical model behind \eqref{eq-Velander-quantile} and \eqref{eq-Velander-quantile-aggr} that is capable of explaining the aforementioned rightward shift and reducing the number of parameters with a parametric representation of \(\beta_\tau\).

\section*{Acknowledgment}

The authors thank Han La Poutré and Werner van Westering for their helpful comments. S.\ S.\ thanks Eric Cator for his valuable mentorship.

\bibliographystyle{IEEEtran}
\bibliography{research-1.bib}{}

\end{document}